\documentclass[aps,prb,reprint,showpacs, superscriptaddress]{revtex4-1}

\usepackage{graphicx}
\usepackage{amssymb}
\usepackage{bm}

\begin{document}

\title{Cryomagnetic scanning tunneling spectroscopy study of multi-gap spectra in superconducting 2$H$-NbSe$_2$}

\author{I. Fridman}
\affiliation{Department of Physics, University of Toronto, 60 St. George St., Toronto, ON  M5S1A7, Canada}
\author{C. Kloc}
\affiliation{School of Materials Science and Engineering, Nanyang Technological University, Singapore}
\author{J. Y.T. Wei}
\affiliation{Department of Physics, University of Toronto, 60 St. George St., Toronto, ON  M5S1A7, Canada}
\affiliation{Canadian Institute for Advanced Research, Toronto, ON, M5G1Z8 Canada}

\begin{abstract}
Scanning tunneling spectroscopy was performed on single crystals of superconducting 2$H$-NbSe$_2$, at 300 mK and in a magnetic field, up to 5 T, applied parallel to the $ab$-plane.  This novel field geometry allows the quasiparticle density-of-states spectrum to be measured under finite superfluid momentum, while avoiding contributions from the vortex-core bound states.  At zero field, we observed a fully-gapped conductance spectrum with both gap-edge peaks and sub-gap kinks.  These spectral features show a systematic evolution with the applied field: the kinks close in while the peaks move apart in low fields, and the zero-bias conductance has a two-sloped behavior over the entire field range, though dipping anomalously at $\sim$ 0.7 T.  Our data was analyzed with recent theoretical models for quasiparticle tunneling into a current-carrying superconductor, and yielded distinct evidence for multiple superconducting gaps coming from various Fermi-surface sheets of different topologies, as well as possible implications on the origin of the coexisting charge-density-wave order.
\end{abstract}

\pacs{71.45.Lr,74.25.Jb,74.55.+v,74.70.Xa}

\maketitle

\section{Introduction}
\label{}

The layered dichalcogenide 2$H$-NbSe$_2$ is a heavily studied material, well known for the coexistence of superconductivity and charge density wave (CDW) order, the former occurring below $T_{c}$ $\approx$ 7.2 K and the latter below $T_{CDW}$ $\approx$ 33 K.  Recent interest has been in the complex band structure of 2$H$-NbSe$_2$, particularly in the role that it plays in both the superconducting and CDW order \cite{Johannes06}.  Angle-resolved photoemission (ARPES) and de Haas-van Alphen measurements have revealed a Fermi surface (FS) with multiple sheets of different topologies located at the center and corners of the Brillouin zone \cite{Yokoya01, Corcoran94}.  Low-temperature ARPES data have shown all four of the either 2D or quasi-2D FS sheets to be gapped below $T_c$, with various superconducting gap amplitudes and anisotropies, while the lone 3D FS sheet at the zone center appears to be ungapped down to 5 K.  Temperature-dependent ARPES data have also implicated various FS regions as possible pockets for CDW nesting, although the precise location of these FS pockets and the exact mechanism responsible for the CDW order are still under debate \cite{Kiss07, Borisenko08, Borisenko09, Feng08}.

Scanning tunneling spectroscopy (STS) is a powerful microscopic probe for studying both superconducting and CDW condensates in complex materials.  For 2$H$-NbSe$_2$, STS has allowed real-space imaging of the CDW surface modulations below $T_{CDW}$, while giving spectroscopic evidence for fully-gapped superconductivity below $T_c$.  The application of STS in a $c$-axis magnetic field has also allowed imaging of the superconducting vortex lattice, by mapping out spatial variations in the tunneling density of states (DOS) \cite{Hess89}.  In these earlier STS measurements, sub-gap structures were seen in the DOS spectra, and attributed either to superconducting gap anisotropy outside the vortex core \cite{Hess90, Machida96} or to pair-potential bound states inside the vortex core \cite{Caroli64}.  More recent STS measurements at ultra low temperatures have indicated that the sub-gap features outside the vortex cores are due to a second superconducting energy gap \cite{Guillamon08}.  This interpretation suggests that 2$H$-NbSe$_2$ is a multiband superconductor, with corroborating evidence from heat-capacity and thermal-transport measurements \cite{Huang07, Boaknin03, Sologubenko03} as well as muon-spin rotation measurements of the vortex-core size \cite{Sonier05}.

A timely challenge for the study of 2$H$-NbSe$_2$ is to reconcile the STS data with the ARPES data.  In particular, it would be crucial to explain how the multiple FS sheets observed by ARPES, showing significant gap anisotropy, could give rise to the fully-gapped quasiparticle DOS spectrum observed by STS.  In this paper we report new STS measurements taken at 300 mK, where we have applied an \emph{in-plane} magnetic field of up to 5 T that generates a screening current on the $ab$-surface.  This novel STS geometry has the advantage of avoiding vortex-core states, and allows a measurement of the spectral dependence on superfluid momentum without complications due to radial and angular dependences of vortical currents \cite{Hess90, Kohen06}.  We observed a systematic spectral evolution with the applied field, with the gap-edge peaks shifting to higher energies and sub-gap kinks shifting to lower energies in low fields.  Furthermore, the zero-bias conductance showed a two-slope behavior over the entire field range, with a dip anomaly at $\sim$ 0.7 T.  Our data was analyzed with recent theoretical models for quasiparticle tunneling into a current-carrying superconductor \cite{Lukic07, Zhang04}, to yield distinct evidence for multiple superconducting gaps, with the pairing involving several FS sheets of various topologies.  Our results also have possible implications on the location of FS pockets responsible for the coexisting CDW order.

\section{Experimental Details}

The home-built cryomagnetic STS apparatus used in this experiment was specially designed for the magnetic field to be applied parallel, rather than perpendicular, to the sample surface.  For tunneling into the $c$-axis face of a superconductor, this field orientation enables a screening current to be generated across its $ab$-surface (see inset of Figure \ref{fig2}).  The Pt-Ir tips used were conditioned using field emission to ensure their robustness, and RF-filters were used throughout the wiring to remove rectification noise.  The tip scanning rate was minimized to avoid heating due to the piezo motion, and the magnetic field was swept slowly to prevent field drifts of the junction.  The $dI/dV$ conductance spectra were acquired using lock-in amplification with the sample biased relative to the tip.  These experimental refinements enabled us to consistently acquire atomically-resolved topographic images and also to achieve maximum energy resolution down to 300 mK and up to 9 T.  The single crystals of 2$H$-NbSe$_2$ used in our experiment were grown by an iodine vapor transport technique \cite{Oglesby94}, and had $T_{c}$ $\approx$ 7.2 K, and upper critical fields $H_{c2}^{\bot}$ $\approx$ 5 T and $H_{c2}^{\|}$ $\approx$ 15 T respectively for field perpendicular to and parallel to the $ab$-plane.  The thick platelet crystals had large $ab$-plane surfaces, which were cleaved just before cooldown.

\section{Results and Discussion}

\begin{figure}
 \includegraphics[width=8.5cm]{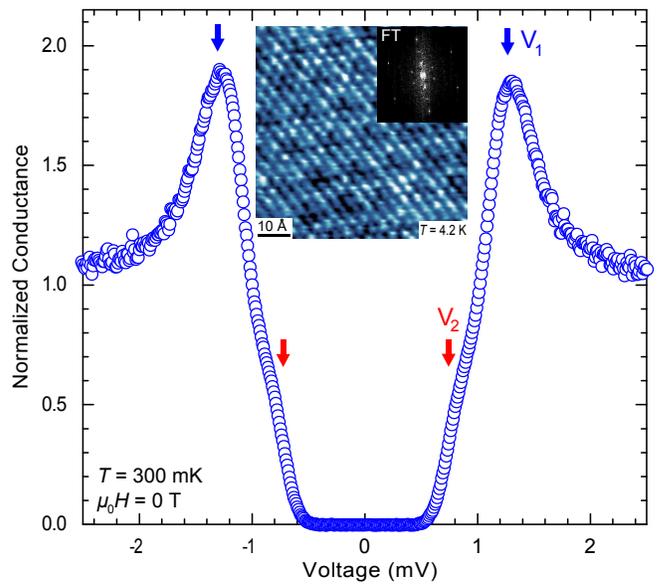}
 \caption{Normalized $dI/dV$ spectrum taken on a 2$H$-NbSe$_2$ crystal at 300 mK in zero magnetic field, showing 
a fully-developed gap structure with gap-edge peaks at $V_1 \approx \pm$1.3 mV and sub-gap features at $V_2 \approx \pm$0.7 mV. The conductance at bias between $\approx \pm$0.6 mV is $\approx$ 0, indicating a fully-gapped DOS.  Inset is an atomically-resolved STM topographic image (60 x 60 $\AA^2$, 1 mV bias, 0.6 nA current, $T$ = 4.2 K), showing the CDW modulations, also seen clearly in the corresponding Fourier transform (FT) image.} \label{fig1}
 \end{figure}

\begin{figure}[t]
 \includegraphics[width=8.5cm]{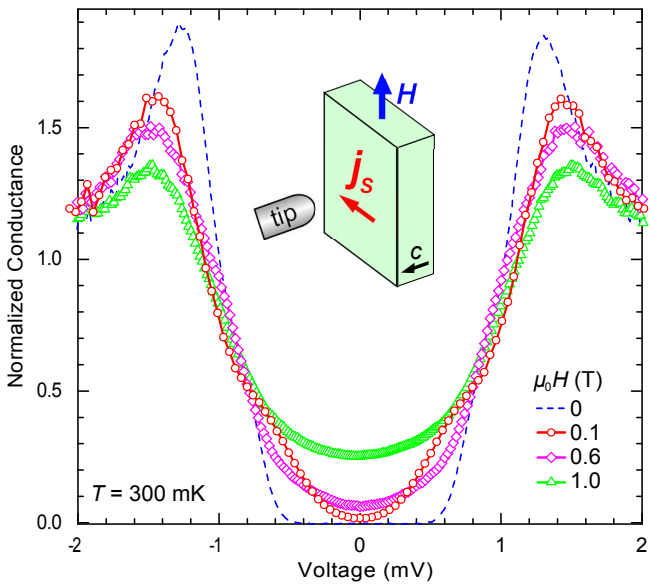}
 \caption{Normalized $dI/dV$ spectra taken on a 2$H$-NbSe$_2$ crystal at 300 mK, at magnetic fields of up to $\mu_{0}H$ = 1.0 T, with zero-field data for comparison (dashed line).  Inset: schematic showing the experimental geometry with the STM tip tunneling into the $c$-axis of the sample, while the in-plane magnetic field produces a screening current ($j_{s}$) on the $ab$-surface.}  \label{fig2}
 \end{figure}

\begin{figure}[ht]
\centering
 \includegraphics[width=8.5cm]{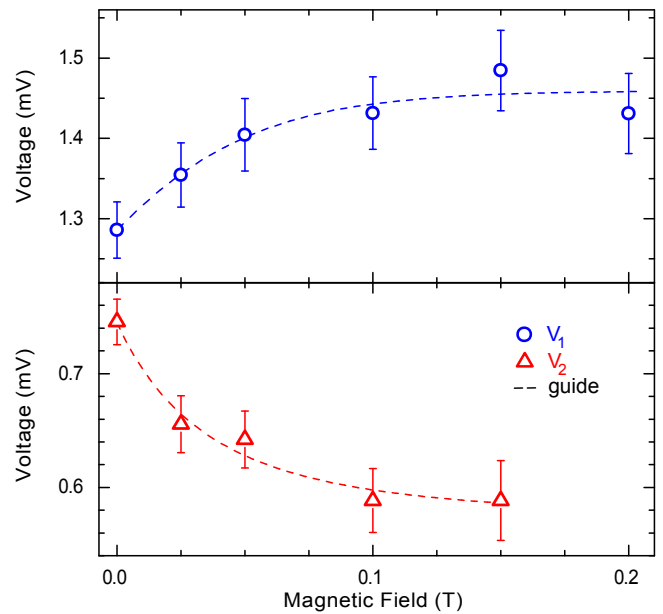}
 \caption{Spectral evolution versus magnetic field up to 0.2 T, showing the field dependences of the gap-edge peaks $V_1$ and sub-gap kinks $V_2$.  As the field is increased, $V_1$ increases from $\approx$ $\pm$1.3 mV  to $\approx$ $\pm$1.45 mV, while $V_2$ decreases from $\approx$ $\pm$0.75 mV to $\approx$ $\pm$0.6 mV and becomes indistinct above $\sim$ 0.2 T.} \label{fig3}
 \end{figure}

Figure \ref{fig1} shows a $dI/dV$ spectrum taken on a 2$H$-NbSe$_2$ crystal at 300 mK in zero magnetic field.  A scanning tunneling microscope (STM) topographic image showing both the atomic lattice and CDW modulations is displayed in the inset.  In the spectrum, which was normalized by the background $dI/dV$ at 4 mV, a fully-developed gap structure is seen, with gap-edge peaks $V_1$ at $\approx \pm$1.3 mV and noticeable sub-gap kinks $V_2$ at $\approx \pm$0.7 mV.  Figure \ref{fig2} shows how the spectrum evolves as a function of the applied magnetic field $\mu_{0}H$.  As the field is increased, $V_1$ shift to higher energies, while $V_2$ shift down to lower energies, as the zero-bias $dI/dV$ steadily rises from zero.  Figure \ref{fig3} more clearly illustrates this spectral evolution, by plotting $V_1$ and $V_2$ \emph{vs.} field up to 0.2 T, when the sub-gap kinks become less discernible.

To understand the field evolution of our tunneling $dI/dV$ spectra, we consider the recent theoretical model of \citet{Lukic07}, which calculates the DOS spectrum for quasiparticle tunneling into a multiband superconductor which is carrying a supercurrent.  The supercurrent adds a Doppler term $\bm{v}_F \cdot \bm{q}_s$ to the quasiparticle energy dispersion $E_k$, where $\bm{v}_F$ is the Fermi velocity and $\bm{q}_s$ is the superfluid momentum, thus modifying the quasiparticle tunneling spectrum \cite{Fulde69}.  For a single-band $s$-wave superconductor, the prior theoretical work by Zhang $\emph{et al.}$ \cite{Zhang04} has shown that the gap edges in the DOS spectrum would first move outward with increasing $\bm{q}_s$, due to Doppler redistribution of the gap amplitude along the FS.  At sufficiently high $\bm{q}_s$, finite-momentum depairing becomes severe enough that the supercurrent density then starts to decrease, causing the gap edges to move inward.  For a two-band $s$-wave superconductor, because the two gaps are depaired at different rates by the supercurrent, their respective features in the DOS spectrum are expected to evolve differently with $\bm{q}_s$.  In our STS data up to 0.2 T, the opposite movements of $V_1$ and $V_2$ with field strongly suggest that they are due to two different energy gaps.  Namely, a small gap which is Doppler-breached and thus collapsing at low fields, along with a larger gap which persists to higher fields.  In fact, the early field breaching of the smaller gap is consistent with the fast rise of the zero-bias $dI/dV$ with field, as seen in Figure \ref{fig3}.

\begin{figure}
 \includegraphics[width=8.5cm]{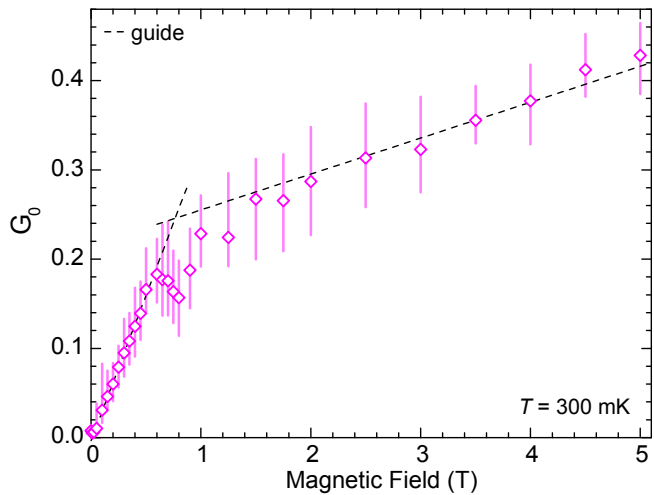}
 \caption{Plot of the normalized zero-bias conductance, $G_0$, versus in-plane magnetic field at 300 mK.  Data were measured at different positions on the surface and averaged to yield the mean (diamonds) and the range between maxima and minima (bars).  A two-slope field dependence is clearly seen in this plot, along with an anomalous dip structure occurring at $\sim$ 0.7 T.} \label{fig4}
 \end{figure}

As further evidence for this two-gap scenario, we also analyze how the zero-bias $dI/dV$ evolves over the entire field range of our measurement.  Figure \ref{fig4} plots the normalized zero-bias conductance, $G_0$, measured at 300 mK, versus the applied field which was gradually increased to 5 T.  At each field value, measurements were made at different positions on the atomically-flat surface and averaged to yield the mean and the range between maxima and minima, indicated by the diamonds and bars respectively in Figure \ref{fig4}.  A two-slope field dependence is clearly seen in this plot, along with an anomalous dip structure occurring at $\sim$ 0.7 T, as featured in the inset.  Similar two-slope field dependences have also been seen for the heat-capacity coefficient $\gamma$ measured in both 2$H$-NbSe$_2$ and MgB$_2$, and interpreted as distinct evidence of two-gap superconductivity in these systems \cite{Huang07, Boaknin03, Bouquet02}.  According to the model of \citet{Lukic07}, since the zero-bias  $dI/dV$ is a measure of the quasiparticle DOS at the Fermi level, the slope change occurs when the smaller of the two gaps begins to close on parts of the FS due to finite-momentum depairing.  The dip anomaly, on the other hand, has also been observed in field-dependent thermal-transport measurements in 2$H$-NbSe$_2$, though its physical origin has not yet been clearly identified \cite{Sologubenko03}.  It should be noted that the dip anomaly appears to coincide with the slope change, thus it is most likely also related to the collapse of the smaller gap.  It is also worth noting that the upper slope in Figure \ref{fig4} undershoots unity when extrapolated to $H_{c2}^{\|}$ $\approx$ 15 T, consistent with there being an additional slope jump between 5 T and 15 T, as the larger gap also becomes Doppler-breached at the higher $\bm{q}_s$. 

The two-gap interpretation of our STS data has important implications for the scenario of multiband pairing in 2$H$-NbSe$_2$.  First, the early collapse of the smaller gap with field, well below $H_{c2}^{\|}$ $\approx$ 15 T, suggests that it comes from the 3D pancake-shaped FS sheet around the Brillouin zone center $\Gamma$, since an in-plane field cannot as effectively induce depairing currents in the other four FS sheets, which are barrel-shaped and have relatively small dispersions along $k_z$ \cite{Johannes06}.  Second, the small size of this gap at 300 mK indicates that it should also be weak against temperature, thus explaining why ARPES was not able to resolve any gap on the pancake FS sheet at 5 K \cite{Kiss07}.  Third, although a total of five FS sheets have been observed by ARPES in the six-fold symmetric Brillouin zone, the two outer FS barrels around $\Gamma$ and $K$ are ostensibly 2D and thus cannot robustly couple with the tunneling matrix element along $k_z$ \cite{Wei98}.  Of the remaining three FS sheets, one is the 3D pancake already discussed, and the other two are quasi-2D barrels with similar superconducting gap maxima, as measured by ARPES \cite{Kiss07}.  This variety of FS topologies could plausibly explain why only two gap structures are manifested in the $c$-axis tunneling spectrum at 300 mK, even though all five FS sheets are evidently involved in the pairing for 2$H$-NbSe$_2$.  

Despite this plausible scenario of multiband pairing in 2$H$-NbSe$_2$, there remains a discrepancy between the STS and ARPES data over the superconducting gap topology.  According to the APRES data of Kiss $\emph{et al.}$ \cite{Kiss07}, the inner quasi-2D barrel around $K$ has a fairly large gap anisotropy, ranging between 1.1 meV and 0.3 meV.  Since this sheet has finite $k_z$-dispersion and is thus expected to contribute to $c$-axis tunneling, it is rather puzzling why the STS spectrum shows a vanishing DOS up to $\sim$ 0.6 meV.  One possible explanation is that the ARPES data, which was taken at 5 K, has enough thermal smearing to produce finite low-energy DOS not seen in the STS data, which was taken at 300 mK.  A more interesting possibility involves $k$-space truncation of the superconducting gap anisotropy by the CDW order \cite{Seo07}.  Namely, pockets on the FS which are connected by the CDW wave vectors are effectively removed from contributing to the quasiparticle DOS spectrum, thus keeping the gap anisotropy from fully showing up in STS.  For this scenario to explain the STS data, the CDW pockets would have to coincide with superconducting gap minima on the FS, $\emph{i.e.}$ located at where the $K-M$ line intersects the inner FS barrels around $K$, as in the scheme proposed by \citet{Borisenko09}.  Both of these possibilities could explain the apparent discrepancy between the STS and APRES data.  In fact, the latter possibility could help to corroborate the location of the CDW pockets, which is still under serious debate, and thus to identify the mechanism responsible for the CDW order in 2$H$-NbSe$_2$.  To elucidate these issues we will carry out a more detailed analysis of our STS data, using the multiband tunneling formalism described above and explicitly accounting for the FS topology observed by ARPES. 

\section{Conclusion}

In summary, we have performed scanning tunneling spectroscopy on single crystals of superconducting 2$H$-NbSe$_2$, at 300 mK and in a magnetic field, up to 5 T, applied parallel to the $ab$-plane.  This novel field geometry effectively avoids vortex-core states, while allowing the quasiparticle DOS spectrum to be measured under finite superfluid momentum.  We observed a fully-gapped conductance spectrum showing both gap-edge peaks and sub-gap kinks, which evolved systematically with the applied field.  Our data was analyzed using recent theoretical models for quasiparticle tunneling into a current-carrying superconductor, and showed the spectral evolution to bear distinct signatures of two superconducting gaps.  Our results indicate that 2$H$-NbSe$_2$ is an ostensibly multiband superconductor, with the pairing process involving several Fermi-surface sheets of different topologies.  An apparent discrepancy between the STS and ARPES data, in relation to the superconducting gap topology, remains to be clarified and may have direct implications on the location of the FS pockets responsible for the coexisting CDW order.

\begin{acknowledgments}
This work was supported by NSERC, CFI/OIT, OCE and the Canadian Institute for Advanced Research under the Quantum Materials Program.
\end{acknowledgments}


%

\end{document}